# 무선통신구간별 전송 데이터 특성을 고려한 드론 보안기준 강화 방안

김대건*, 이현호**

Ⅰ. 서론
Ⅱ. 드론 관련 보안기준
Ⅲ. 드론 구성, 무선구간 통신 방식 및 전송 데이터 특성
Ⅳ. 드론 무선구간 보안취약점의 이론적 모델
Ⅴ. 보안기준 강화 방안
Ⅵ. 결론

### 요약

4차 산업혁명 기술의 발전에 따라 드론의 종류 및 활용 범위가 급격히 증가하고 있다. 드론에 대한 관심이 증대되는 만큼 관련된 보안취약점에 대한 연구도 활발히 이루어지고 있으며, 이를 예방하기 위한 보안대책에 대한 논의도 함께 발전되어 왔다. 우리 군에서도 이에 대한 다양한 보안대책을 수립하여 적용하고 있으며, 특히 드론의 무선 전송구간 데이터를 보호하기 위한 방책으로 KCMVP 인증 암호모듈을 탑재한 드론을 도입하도록 하고 있다. 그러나 이와 같이 인증된 암호모듈을 장착하고 있음에도 불구하고 드론의 운영환경 및 구조적 특성으로 인해 여러 보안취약점에 노출될 가능성이 존재한다. 본 논문에서는 이러한 보안취약점에 대한 이론 모델과 이를 보완하기 위한 방안을 제시하였다.

핵심어 : 드론, 무선통신, 보안취약점, KCMVP, 보안기준

# Ⅰ. 서론

4차 산업혁명 기술의 발전에 따라 UAV[1]로 지칭되던 종류의 비행체가 드론의 범주로 통칭되는 등 그 종류 및 활용 범위가 급속히 증대되고 있다. 민간, 공공, 군사 등 여러 분야에서는 각 기능을 수행하기 위한 다양한 활용방안이 추진되고 있으며, 이를 구현하기 위해 요구되는 영상처리 기술[2] 등 각종 기반 기술이 개념적으로 제시되고 또 구현되고 있다.

드론에 대한 관심이 증대되는 만큼 보안취약점에 대한 연구도 활발히 이루어지고 있다. 김대건[3]은 4개 제조사의 상용 드론의 조종용 모바일 앱을 분석하여 암호키 노출 및 중요 정보 평문 전송 등 18개의 보안취약점을 식별 후 이를 5개 유형으로 분류하였다. Giray[4]는 드론의 GPS 신호를 스푸핑하여 드론의 제거 권한을 탈취하는 방법을 제시하였고, Yaacoub[5]은 서베이 논문에서 다양한 드론 공격 유형 및 시나리오를 제시하였다. Rodday[6]는 블랙햇 아시아 2016에서 드론 제어기의 프로토콜을 분석하여 중간자 공격을 수행하는 기법을 발표하였다. Hartmann[7]은 드론이 운용되고 드론 거부작전이 시행된 분쟁 사례를 제시하면서 향후 이러한 능력이 사이버 도메인을 지배하기 위한 새로운 영역이 될 것이라

---

[1] Unmaned Aerial Vehicle
[2] 김대건, 백승원, 백승수, "미래지향적 지상군 소부대 무인항공기(UAV)에 대한 영상처리기술 기반의 요구기능 제안 및 운용개념 검증." 한국군사학논집 71권, 1호 (2015)
[3] Daegeon Kim, Huy Kang Kim, "Security Requirements of Commercial Drones for Public Authorities by Vulnerability Analysis of Applications." arXiv (2019)
[4] Sait Murat Giray, "Anatomy Of Unmanned Aerial Vehicle Hijacking With Signal Spoofing." 6th International Conference on Recent Advances in Space Technologies (2013)
[5] Jean-Paul Yaacoub, Hassan Noura, Ola Salman, Ali Chehab, "Security analysis of drones systems: Attacks, limitations, and recommendations." Internet of Things Vol. 11 (2020)
[6] Nils Rodday, "Hacking a Professional Drone." Black hat ASIA (2016)
[7] Kim Hartmann, Keir Giles, "UAV Exploitation: A New Domain for Cyber Power." 8th International Conference on Cyber Conflict Cyber Power (2016)

고 하였다.

이러한 보안취약점들로 인해 국가 및 공공기관, 국가중요시설 등 공공 분야에서는 요구되는 보안기준을 충족하는 드론이 도입되도록 하기 위한 보안기준을 마련하고 있는 추세이다. 특히, 대다수 보안기준에서는 무선통신구간의 데이터를 보호하기 위해서 암호화를 적용하도록 하고 있는데 제시된 기준 만으로는 의도한 보안 수준을 충족하지 못하는 상황이 발생할 수 있다. 본 논문에서는 보안기준에도 불구하고 이렇듯 취약점이 지속될 수밖에 없는 환경 조건을 암호 이론을 적용하여 분석하고 보완 방안을 제시한다.

이를 위한 논문의 구성은 다음과 같다. Ⅱ장에서 드론과 관련된 여러 보안기준을 소개하고 Ⅲ장에서는 드론의 구성 및 무선구간 통신에 적용되는 다양한 방식과 무선구간별 전송되는 데이터의 특성을 알아본다. Ⅳ장에서는 드론의 전송구간 통신방식의 특성, 운용환경 및 구성특징 등으로 인해 보안기준을 적용하더라도 발생하는 보안취약점에 대한 이론적 모델을 제시한다. Ⅴ장에서는 이러한 취약점을 해소하기 위한 다양한 방안을 제시하고, Ⅵ장에서 향후 연구과제와 함께 논문을 결론짓는다.

## Ⅱ. 드론 관련 보안기준

드론의 안전한 활용을 위해 요구되는 보안에 대한 논의들이 많이 이루어지고 있는데, 그 중에서도 보안 기능 및 보안 요구사항에 대한 공식적인 기준 또는 기술표준 문서들을 살펴볼 수 있다.

우선, TTAK.KO-12.0317과 IoTFS-0079는 유사한 형태와 내용으로 구성되어 있는데, 드론 기반 서비스 시스템을 6개 구성요소(서비스 요청자, 서비스 제공 조직, 지상 관제사, 드론, 구역 관리 장치)로 구분하여 각 구

성요소의 보안 요구사항과 구성요소 간 인터페이스의 보안 요구사항을 구분하여 제시하고 있다.

KISA-GD-2019-0005와 CLP.11은 직접적으로 드론과 관련된 보안기준 문서는 아니지만 다양한 IoT 제품에 범용적으로 적용할 수 있도록 구성된 만큼 드론에도 적용이 가능하다. 특히, CLP.11은 IoT 제품에 대한 일반적인 기준을 드론에 적용한 예를 제시하고 있다.

한국인터넷진흥원에서 2020년에 발간한 '드론 사이버보안 가이드'에는 드론 서비스를 구성하는 드론(구동부, 제어부, 페이로드, 통신부 등)과 주요 시스템(드론, 지상제어장치, 정보제공장치 등)에 예상되는 보안 위협 시나리오와 보안 요구사항이 제시되어있다[8]. 해당 가이드에서는 드론 무선 구간의 보안 위협을 '중간자 공격' 및 '데이터 손실'의 위협 시나리오로 구분하여 대응방안을 제시하였으나, 본 논문에서 제시한 보안위협 분석 및 무선 구간별 고려사항이 일부 간과된 측면이 있다.

ISO 21384-3은 드론의 기능 자체보다는 운용 절차에 관한 국제 표준으로, 안전한 비행과 운항 간 사고 예방을 위해 관리·운용 측면에서 요구되는 사항들을 개략적인 수준에서 다루고 있다[9].

이들 보안기준에서는 디지털 매체 등 데이터를 적절하게 보호하도록 명시하고 있는데, 그 중에서도 TTAK.KO-12.0317, IoTFS-0079, KISA-GD-2019-0005는 전송구간의 중요 데이터에 대한 기밀성을 보장하도록 하고 있다. 특히 KISA-GD-2019-0005는 중요 데이터를 전송 시 안전한 암호 알고리즘을 사용하고, 무선통신구간에서 알려진 프로토콜을 사용해야 할 경우 신뢰된 보안 프로토콜을 사용하도록 정하고 있다.

이상의 보안기준 및 기술표준 문서들을 정리하면 <표 1>과 같다.

---

[8] 국토교통부 보도자료, "과기정통부-국토부, 안전한 융합서비스 구현을 위한 드론 사이버보안 가이드 발표." (2020. 12. 1)
[9] ISO. *ISO 21384-3:2019(en) Unmanned aircraft systems － Part 3: Operational Procedures* (Geneva : International Organization for Standardization, 2019), p. vi.

<표 1> 드론 관련 국내·외 보안기준

| 기준 번호 | 기준 명칭 | 발간 기관 | 발간 년도 |
|---|---|---|---|
| TTAK.KO -12.0317 | 드론 기반 서비스를 위한 보안 요구사항 | 한국정보통신 기술협회 | 2016 |
| IoTFS-0079 | 드론 기반 사물인터넷 서비스를 위한 보안 요구사항 | 사물인터넷포럼 | 2015 |
| KISA-GD -2019-0005 | 사물인터넷(IoT) 보안 시험·인증 기준 해설서 | 한국인터넷 진흥원 | 2019 |
| - | 드론 사이버보안 가이드 | 한국인터넷 진흥원 | 2020 |
| CLP.11 (ver. 2.2) | IoT Security Guidelines Overview Document | GSMA[10] | 2020 |
| ISO 21384-3 | Unmanned aircraft systems - Part 3: Operational procedures | ISO[11] | 2019 |

이외, 각 국에서는 특히 국가보안 측면에서 고도의 보안이 요구되는 용도로 상용 드론 활용 시 디지털 매체 등 데이터를 보호하기 위한 내부 기준을 마련했거나 정책적으로 통제하고 있다. 미국의 경우는 2019년 국토안보부(Department of Homeland Security, DHS) 산하 사이버보안 및 기반시설보안청(Cybersecurity & Infrastructure Security Agency, CISA)에서 상용 드론을 통해 처리·저장되는 데이터의 기밀성·무결성을 위한 방편으로 암호화를 적용하도록 하였고, 그 기준은 '단말기를 위한 저장소 암호화 기술 가이드(Guide to Storage Encryption Technologies for End User Devices)'를 따르도록 하였다[12]. 일본은 2021년부터 정부에서 도입하는 드론에 대해 운항 기록과 처리·저장된 자료를 외부의 탈취로부터 보호하기 위해 내각관방에 의한 보안성 심사 절차를 마련하여

---

10) Global System for Mobile communications Association, 세계이동통신사업자연합회
11) International Organization for Standardization, 국제표준화기구
12) DHS CISA, "Cybersecurity Best Practices for Operating Commercial Unmanned Aircraft Systems," https://www.cisa.gov/sites/default/files/publications/CISA%20Cybersecurity%20Best%20Practices%20for%20Operating%20Commerical%20UAS%20%28508%29.pdf (검색일: 2021. 3. 25).

통제하고 있다13). 우리나라 또한 군에서 상용 드론을 활용하여 군사비밀이 아니더라도 비공개 업무자료 등을 촬영·처리하고자 할 경우 '국방보안업무훈령' 등 보안관계 규정에서 KCMVP14) 인증을 받은 암호모듈을 탑재하는 등의 보안대책 강구를 통해 무선 구간에서 전송·처리되는 데이터를 안전하게 보호할 수 있는 제품을 도입하도록 규정하고 있다.

## Ⅲ. 드론 구성, 무선구간 통신 방식 및 전송 데이터 특성

### 1. 드론의 일반적인 구성

드론의 구성체는 다양한 방식으로 구분할 수 있으나 크게 비행체와 지상통제장비로 구분할 수 있다. 이 중 지상통제장비는 <그림 1>과 같이 Rodday15)가 구분한 것처럼 임무계획장치, 원격제어기, 원격지시기 등으로 세분화 될 수 있다. 이러한 장비 구성의 구분은 정형화 되어있지 않으며 기능적 측면에서 여러 장비로 구분하여 구성 될 수도, 하나의 장비로 통합될 수도 있다.

<그림 1>에서 보는 바와 같이 드론이 구성된다고 가정할 때 무선통신은 '원격제어기-비행체', '원격지시기-비행체', '원격지시기-임무계획장치' 구간에서 이루어진다. 무선 데이터의 전송 방식 및 프로토콜은 각 구간별 상이하게 적용될 수 있고, 전송 방향도 무선구간별로 양방향, 또는 단방향(상·하향) 링크로 상이하게 구성될 수 있다.

---

13) 每日新聞, "政府機関、中国製ドローン新規購入を排除　情報漏えい・乗っ取り防止を義務化," https://mainichi.jp/articles/20200925/k00/00m/040/245000c (검색일: 2021. 3. 25).
14) Korea Cryptographic Module Validation Program, 한국형 암호모듈검증
15) Nils Rodday, "Hacking a Professional Drone." Black hat ASIA (2016), p.5 재구성

<그림 1> 드론의 일반적인 구성

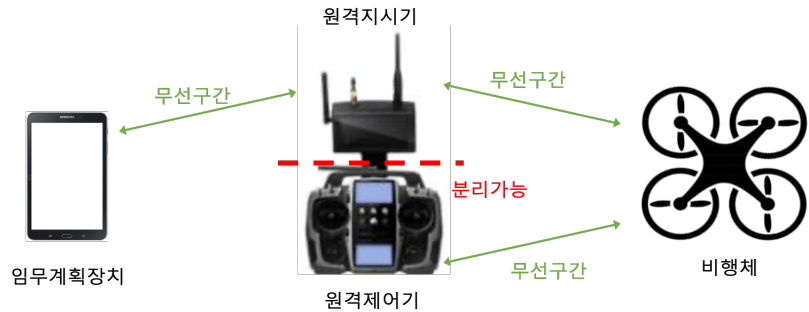

<표 2> 드론 구성 요소별 기능

| 구성 요소 | 기 능 |
|---|---|
| 임무계획장치 | 비행체의 임무를 계획하고 각종 지원 및 운용 상황실과 같은 기능을 수행 |
| 원격지시기 | 비행체와 기타 구성요소 사이에서 통신을 중계 |
| 원격제어기 | 비행체의 이·착륙 및 임무지역에서의 비행체 제어를 수행하여 임무 수행을 통제. 운용 환경에 따라 이·착륙을 제어하는 이착륙통제소와 임무지역에서의 비행을 통제하는 지상통제소로 구분하여 구성 될 수도 있음. |
| 비행체 | 원격제어기의 통제 또는 미리 정의된 내용에 따라 장착된 임무센서를 활용한 신호 수집 및 물자 수송 등의 임무 수행 |

## 2. 무선구간 통신 방식 및 프로토콜

### 가. 통신 방식

드론의 무선구간은 현존하는 다양한 무선통신기술이 적용될 수 있다. Neji[16])는 다양한 드론 무선통신기술을 통달거리, 에너지 소비, 통신속도,

---

16) Najett Neji, Tumader Mostfa, "Communication technology for Unmanned Aerial Vehicles: a qualitative assessment and application to Precision Agriculture."

보안레벨 등의 관점에서 비교하였는데 이를 <표 3>과 같이 정리할 수 있으며, 이밖에도 주로 900MHz, 1.3GHz, 2.4 GHz 및 5.8 GHz 대역의 일반적인 아날로그 무선통신(Radio Frequency)을 사용하기도 한다.[17]

<표 3> 드론의 무선통신 방식 비교

| 통신 방식 | 통달거리 | 에너지 소비 | 통신속도 | 보안레벨 | 반응속도 |
|---|---|---|---|---|---|
| Wi-Fi | 50-250 m | 835 Mw | 6,5-200 Mbps | 중간 | 50 ms |
| Wimax | 5km | 3,2 W | 1 Gbps | 높음 | 25-40ms |
| LTE-U | 2-5 km | 1W | 1 Gbps | 높음 | 9ms |
| Zigbee | 500 m | 36,9 mW | 250 Kbps | 높음 | 20 ms |
| Bluetooth | 10m > | 215 Mw | 1-3 Mbps | 중간 | 100 ms |
| Ingenu | 2km > | 160 Mw | 600 Kbps > | 높음 | insensitive |
| Lora | 2km > | 100 Mw | 100 bps | 높음 | low latency |
| SigFox | 2km | 100 Mw | 50 Kbps | 낮음 (암호화 미지원) | insensitive |
| 5G | 50 km (expected) | N/A | 50 Gbps (expected) | 중간 (expected) | 1ms > |

### 나. 프로토콜

특정 통신방식의 경우 특화된 프로토콜을 사용하기도 하지만 기본적으로 TCP/IP 모델과 유사한 구조를 지니고 있다. 프로토콜이 특정된 통신 방식이 아닐 경우 응용계층에서 HTTP와 같은 웹 방식의 프로토콜이나 드론에 특화시키기 위해 경량화 된 MAVLink[18]와 같은 프로토콜을 사용할 수도 있다. 이와 같이 공개된 프로토콜이 아닌 경우 자체 개발한 프로토콜을 적용하여 제어신호를 생성 후 표준에 정의된 특정 무선통신방식 또는 일반적인 RF 통신을 적용하여 드론을 제어할 수도 있다.

---

International Conference on Unmanned Aircraft Systems (2019)

17) Mike Richardson. "Drones: Detect, identify, intercept and hijack," last modified December 2, 2015, https://www.nccgroup.com/uk/about-us/newsroom-and-events/blogs/2015/december/drones-detect-identify-intercept-and-hijack/

18) Micro Air Vehicle Link

## 3. 드론 무선통신구간별 전송 데이터 특성

드론에 대한 위험을 이해하기 위해서 앞에서와 같이 드론을 구성하는 유형 자산의 특성을 구분하는 것은 물론 무형 자산, 즉 유형 자산 사이에서 전송되는 데이터의 특성을 구분하여 이해하는 것도 중요하다. 앞서 <그림 1>과 같은 드론의 구성에 따라 구분한 무선통신구간별 전송 데이터의 특성은 <표 3>과 같이 요약될 수 있다.

<표 3> 무선통신구간별 전송 데이터 특성

| 통신구간 | | 데이터 유형 | 전송 거리 |
| --- | --- | --- | --- |
| ① | 임무계획장치 ~ 원격지시기 | - 비행체 관련 : 지리 좌표, 고도, 비행 방향 등 | 단거리 |
| ② | 원격지시기 ~ 비행체 | - **임무 관련 : 영상, 신호 등 수집 정보, 표적 정보 등** | 장거리 |
| ③ | 원격제어기 ~ 비행체 | - 비행체 관련 : 지리 좌표, 고도, 비행 방향 등<br>- **임무 관련 : 영상장비 등 센서 제어 정보 등** | 장거리 |

<표 2>에서 원격지시기의 기능을 '비행체와 기타 구성요소 사이에서 통신을 중계'하는 것이라 하였으므로 '임무계획장치~원격지시기'(구간 ①)와 '원격지시기~비행체'(구간 ②) 간의 전송 데이터의 유형은 동일하다. <표 3>의 '데이터 유형'에서와 같이 구간 ①~②와 구간 ③에서는 공통적으로 비행체 제어 및 상태 정보가 유통될 수 있다. 구간 ①~②에서는 비행체의 각종 임무센서에서 수집되는 정보가 유통되고 구간 ③에서는 임무센서의 제어 및 상태 정보가 유통될 수 있는데 비행체의 운용 목적을 고려한 정보의 가치 측면에서 임무센서에서 수집되는 정보의 가치가 더욱 높은 것으로 평가할 수 있다. <표 3>의 '전송 거리'는 절대적인 특성이 아니며 운영 환경에 따른 시스템 구성에 따라 상이 할 수 있다. 다만, 무선통신구간의 보안기준을 고려함에 있어 주목해야 할 점은 각 통신구간별 전송 거리 및 전송 데이터 용량 등에 따라 상이한 무선통신 방식이 사용될 수 있다는 것이다.

# Ⅳ. 드론 무선통신구간 보안취약점의 이론적 모델

본 장에서는 드론의 무선구간 전송 데이터의 보안취약점에 대한 암호학적 모델을 제시한다.

Ⅱ장에서 언급한 바와 같이 드론에 KCMVP 인증 암호모듈을 탑재하도록 보안기준이 설정된 이유는 무선구간 전송 데이터를 안전하게 보호하기 위함이다. 그러나 본 장에서 제시하는 보안취약점의 근본적인 원인은 KCMVP 인증 보안모듈을 탑재하는 보안기준이 충족되었음에도 불구하고 드론의 운영환경 및 구조적 특성으로 인해 보안기준이 설정된 본래의 보안 목표가 달성되지 않을 수 있다는데 있다.

## 1. 드론 운영환경 특성으로 인한 재생 공격(Replay Attack)

드론의 무선 전송 데이터는 공중으로 전파되어 신호의 특성을 알고 있는 제 3자에 의해 탈취될 수 있다. 이러한 위협에 대한 보호 대책으로써 무선 데이터를 암호화하도록 하고 있으나 무선 데이터의 암호화가 적용된 드론의 움직임을 제 3자가 관측할 수 있다는 운영환경의 특성으로 인해 암호화된 무선 데이터에 대한 재생공격이 가능한 위협이 존재한다. 드론의 움직임을 제 3자가 관측할 수 있다는 것은 암호문에 대한 평문이 노출된다는 의미이기 때문이다. 이러한 운영환경의 특성에 의한 보안취약점을 <그림 2>와 같이 도식화될 수 있다.

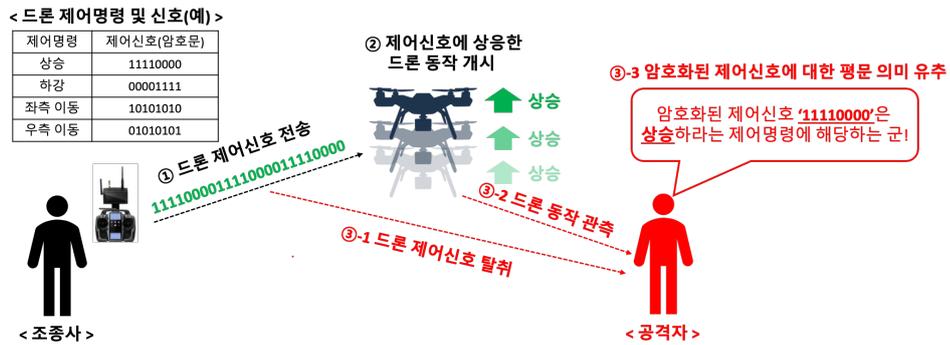

<그림 2> 드론 운영환경 특성에 의한 보안취약점

<그림 2>의 환경을 수식 (2.1)과 같이 표현할 수 있다.

$$\begin{aligned}
&① \ signal \ = \ Cipher(command) \\
&\quad where \ command \ \in \ \{Up, \ Down, \ Fowrard, \ Backward...\}, \\
&\quad when \ command_1 = command_2, \ then \ Cipher(command_1) = Cipher(command_2) \\
&② \ movement \ = \ Action(signal)
\end{aligned} \quad (2.1)$$

즉, 평문에 해당하는 유한한 제어명령(*command*)와 이에 대한 암호문에 해당하는 제어신호(*signal*)가 1:1 대응의 관계일 경우(①) 공격자는 암호문에 해당하는 평문을 직접 확인할 수 있으므로(②) 암호문 재생공격이 가능하다.

공개키 암호의 경우 다양한 운영모드를 제공하고 있는데 이 중에서 ECB모드의 경우 암호문 재생공격에 취약하지만 KCMVP에서 인증하는 운영모드에 포함되어 있다. 또한 RSA-OAEP와 같이 암호화 과정에서 난수를 사용하는 공개키 암호는 암호문 재생공격에 안전하지만 그렇지 않은 일반적인 RSA 암호 알고리즘의 경우 암호문 재생공격에 취약하다.

또한, 제시된 운영환경 특성과 같이 암호문에 대응하는 평문이 노출될

경우 기지 평문 공격(KCA, Known Ciphertext Attack)을 통한 암호키 획득을 시도할 수 있는데 이는 암호문 단독 공격(COA, Ciphertext Only Attack) 등과 같은 기타 암호문 공격방식에 비교하여 복잡도가 낮은 것으로 알려져 있어 상대적으로 취약하다고 할 수 있다.

## 2. 드론의 구조적 특성으로 인한 무선 데이터 암호화 미적용

드론이 <그림 1>과 같이 구성되어 있을 때 원격지시기에만 암호모듈이 장착되어있을 경우 <그림 3>과 같이 암호모듈에 의한 암호화가 적용되지 않는 무선통신구간(원격제어기↔비행체)이 발생하게 된다.

<그림 3> 드론의 구조적 특성에 의한 보안취약점

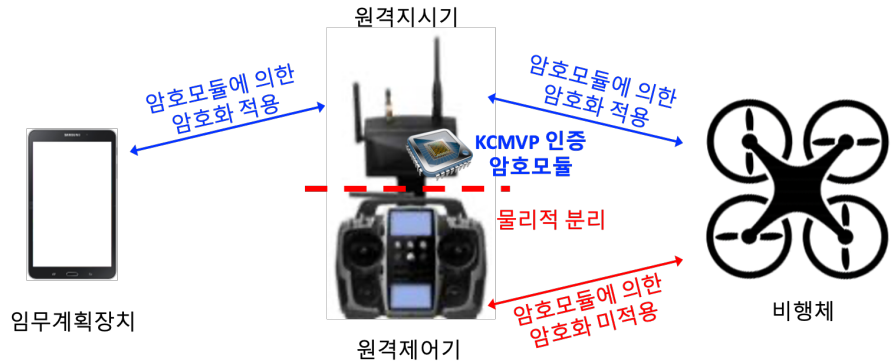

이러한 보안취약점을 수식 (2.2)와 같이 표현할 수 있다.

$$\exists D \text{ where } D \in \{\text{임무계획장치, 원격제어기, 원격지시기, 비행체...}\}, \quad (2.2)$$
$$Signal(D_1, D_2) = Cipher(Signal(D_1, D_2))$$

즉, 드론을 구성하는 특정 장비($D_1$, $D_2$) 사이의 평문 무선 데이터

($Signal(D_1, D_2)$)와 암호화된 무선 데이터($Cipher(Signal(D_1, D_2))$)가 동일하다는 의미이므로 해당 구간에는 암호화가 적용되지 않았다는 것을 뜻한다. 만약, 원격지시기가 임무계획장치 및 비행체와의 통신을 위해 서로 다른 무선통신 방식을 사용한다면 장착된 KCMVP 인증 암호모듈에 의해 두 무선 구간의 전송 데이터가 모두 암호화 되지 않을 수도 있다. 이러할 경우 수식 (2.2)의 취약점이 발현되는 무선 구간이 증가할 것이다.

## Ⅴ. 보안취약점 해소 방안

본 장에서는 앞 장에서 살펴보았던 각각의 보안취약점들에 대해 해소하기 위한 방안을 제시한다.

### 1. 운용 모드, 제어 프로토콜 요구사항 구체화

단순히 드론에 KCMVP 인증 암호모듈이 탑재되었는지 여부만 확인하는 것만으로는 앞 장에서 제시하고 있는 암호문 재생 공격 및 기지 평문 공격에 대한 취약 여부를 확인할 수 없다. 그러므로 탑재된 암호모듈이 어떤 암호 알고리즘을 탑재하고 있으며 어떤 방식의 암호 모드로 동작하는지 살펴볼 필요가 있다.

먼저, 암호모듈이 대칭키 암호를 제공하는 경우 ECB, CFB 등 재생 공격에 취약한 모드의 적용 여부를 확인하여야 한다. 특히 ECB 모드는 간단하면서도 신속하며 암·복호화 모두 병렬처리가 가능하다는 이점이 있지만 재생 공격에 취약하기 때문에 권장될 수 없다. CBC, CTR, GCM 등 재생 공격에 대응 가능한 모드 사용이 권장된다.

공개키 암호를 제공하는 경우 RSA-OAEP와 같이 동일한 평문을 암호화하더라도 암호화 과정에서 난수화 패딩을 적용하여 매번 다른 암호문

을 출력하는 알고리즘이 적용되었는지 등의 여부를 확인하여야 한다.

　암호모듈의 운영모드 설정 확인 이외 무선통신 프로토콜도 살펴볼 수 있다. 이를 위해 프로토콜 설계 시 SRTP(Secure Real-Time Protocol)처럼 시시각각 변하는 Sequence Number 등 재생 공격 보호 논리와 관련된 사항들이 포함되어 있는지 확인할 필요가 있다.

　이상의 내용을 점검 항목 형태로 구성하여 정리하면 <표 4>와 같다.

<표 4> 운용 모드, 제어 프로토콜에 대한 점검 항목

| 점 검 항 목 | | | 확인 |
|---|---|---|---|
| 1. | KCMVP 인증 암호모듈이 탑재되었는가? | | ○ (탑재)　✕ (未탑재) |
| | 1.1. | 암호모듈이 대칭키 암호를 제공하고 있는가? | ○ (제공)　✕ (未제공) |
| | | 1.1.1. 재생 공격에 안전한 암호화 모드를 적용하고 있는가? | ○ (적용)　✕ (未적용) |
| | 1.2. | 암호모듈이 공개키 암호를 제공하고 있는가? | ○ (제공)　✕ (未제공) |
| | | 1.2.1. 난수 사용 알고리즘이 적용되어 있는가? | ○ (적용)　✕ (未적용) |
| 2. | 안전한 무선통신 프로토콜이 사용되었는가? | | ○ (사용)　✕ (未사용) |
| | 2.1. | 재생 공격 보호 관련 논리가 포함되어 있는가? | ○ (포함)　✕ (未포함) |

## 2. 모든 무선 전송구간에 대한 암호화 적용 확인

　Ⅳ.2장에서 제시한 바와 같이 KCMVP 인증 암호모듈의 장착여부가 모든 무선 전송구간 데이터에 대한 암호화 여부를 보증하는 것이 아니다. 따라서 드론 제품에 탑재된 암호모듈의 KCMVP 인증서 유효성 여부만 확인하는 것이 아닌 해당 모든 무선구간 전송 데이터에 해당 암호모듈을 적용하여 암호화하는지 여부를 확인해야 한다. <그림 3>에서는 원격지시기에만 KCMVP 인증 암호모듈이 적용되어 있고 원격제어기에는 적용되

지 않아 원격제어기↔비행체 구간에서 암호화가 적용되지 않는 취약점이 존재했다. 이를 보완하기 위해 <그림 4>와 같이 원격제어기에도 KCMVP 인증 암호모듈을 탑재하는 방안을 강구할 수 있다.

<그림 4> 드론의 구조적 특성에 의한 보안취약점 보완 방안

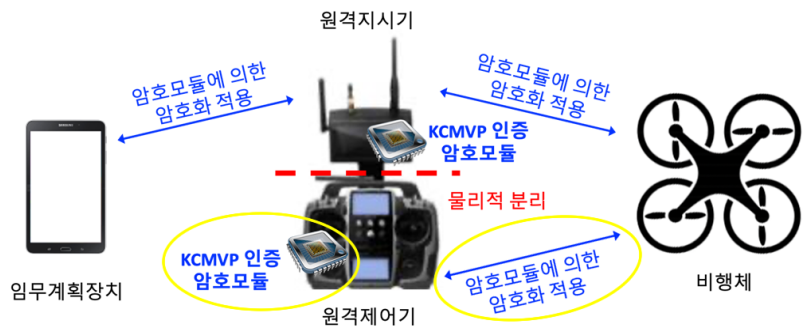

장비 특성상 각 무선 전송구간이 상이한 통신 방식을 사용함에 따라 모든 구간에 KCMVP 인증 암호모듈을 통한 암호화가 제한될 수도 있을 것이다. 이렇게 되면 통신 방식 별 소비전력, 통신거리, 연산처리성능 등 요구사항이 상이할 것이고 <표 3>에서와 같이 무선 구간별 유통되는 데이터의 유형도 달라질 수 있으므로, <표 5>와 같이 무선 구간별 암호화 강도를 차등 적용하는 방안을 모색할 수 있다.

단, 이와 같이 암호화 강도를 차등 적용하는 경우 Ⅲ.3절에서 분석한 바와 같이 가장 고가치 정보라 할 수 있는 임무센서의 수집정보에 가장 높은 암호화 강도를 적용할 필요가 있다.

<표 5> 무선통신구간별 암호화 강도 적용안(예)

| 통신구간 | 거리 | 데이터 유형 | 정보 가치 | 암호화 강도 |
|---|---|---|---|---|
| ① 임무계획장치 ~ 원격지시기 | 단거리 | <비행체 관련> 지리 좌표, 고도, 비행 방향 등 | 고 | 고 |
| ② 원격지시기 ~ 비행체 | 장거리 | <임무 관련> 영상, 신호 등 수집 정보, 표적 정보 등 | | |
| ③ 원격제어기 ~ 비행체 | 장거리 | <비행체 관련> 지리 좌표, 고도, 비행 방향 등 <임무 관련> 영상장비 등 센서 제어 정보 등 | 중 ~ 저 | 중 ~ 저 |

# Ⅵ. 결론

본 논문에서는 기존에 군에서 적용하고 있는 드론에 대한 보안기준에도 불구하고 운영환경 및 구조적 특성으로 인해 여전히 발생 가능한 보안취약점에 대한 이론적 모델과 이에 대한 해소방안을 제시하였다. 이를 위해 먼저, 드론 관련 국내·외 보안 기준과 특히 국방·안보 목적으로 사용되는 경우에 대한 정책적 통제 사례를 살펴보고 데이터 전송·처리에 관하여 정하고 있는 사항들을 확인했다. 다음으로는 드론의 특성, 운용환경 및 구성 특징, 무선통신구간별 전송 데이터 특성을 살펴보고, 기인될 수 있는 두 가지 보안취약점에 대한 이론적 모델을 제시했다. 마지막으로 이들 보안취약점에 대한 해소·대응 방안을 제시하고자 했다.

드론의 효용성을 생각해봤을 때 향후 드론의 군사적 활용 범위가 확대될 것은 의심할 여지가 없다. 특히, 도입 과정 및 기간이나 비용 대비 편익 측면을 고려하면 군 전용으로 연구개발된 드론을 도입하는 경우보다 상용 드론을 도입하는 경우가 많을 것으로 전망된다. 따라서 군에서는 기

술 발전과 운영환경의 변화를 고려하여 드론에 관해 예상 가능한 보안취약점을 지속적으로 식별·분석해야 하며, 이에 부합하는 합리적인 보안기준을 마련하고 정책·제도화하는 등의 노력을 강구해야 한다.

  군 병력 감축 및 구조 개편, 4차 산업혁명은 우리 군에 상당한 도전이 되고 있다. 이를 극복하고 기술 집약적 군으로 발전하기 위해서는 드론과 같이 신기술을 적용한 무기·전력지원체계를 안전하게 도입하는 것이 긴요한만큼, 관련된 보안위험을 적절하게 통제하기 위한 보안대책에 대한 지속적인 연구가 필요하겠다.

# 참고문헌

# Security Enhancement of Drone Considering the Characteristics of Data Transmitted between Wireless Channels

Daegeon Kim[*], Hyeonho Lee[**]

## Abstract

With the development of the fourth industrial revolution technology, the type and scope of use of drones are rapidly increasing. As interest in drones increases, research on related security vulnerabilities is actively being conducted, and discussions on security measures have also been developed to prevent them. Military has established and applied various security measures for this, and in particular, drones equipped with KCMVP-certified cryptographic modules are required to be introduced as a way to protect data on the drone's wireless channels. However, despite being equipped with such a certified cryptographic module, the drone's operating environment and structural properties have the potential to expose it to several security vulnerabilities. In this paper, we present a theoretical model for these security vulnerabilities and measures to complement them.

Key Words: Drone, Security Vulnerabilities, KCMVP, Security Criteria

[*] Officer in Defence Security Support Command, Ph. D. Candidate of Korea University, dgkim@dssc.mil.kr
[**] Instructor in Dept. of Security, Defence Security Support School, cyber@dssc.mil.kr